\documentclass[twocolumn,showpacs,preprintnumbers,amsmath,amssymb]{revtex4}

\usepackage{epsfig}
\usepackage{graphicx}
\usepackage{dcolumn}
\usepackage{bm}

\newcommand{\beq}{\begin{equation}}
\newcommand{\eeq}{\end{equation}}
\newcommand{\beqa}{\begin{eqnarray}}
\newcommand{\eeqa}{\end{eqnarray}}

\newcommand{\bit}{\begin{itemize}}
\newcommand{\eit}{\end{itemize}}
\newcommand{\btab}{\begin{tabular}}
\newcommand{\etab}{\end{tabular}}

\newcommand{\anti}[1]{\overline{#1}}

\newbox\rotbox

\begin{document}

\preprint{ADP-05-05/T615}
\title{A topological model of composite preons}
\author{Sundance O. Bilson-Thompson}
\email{sbilson@physics.adelaide.edu.au}
\affiliation{Centre for the Subatomic Structure of Matter, Department of Physics, 
University of Adelaide, Adelaide SA 5005, Australia}
\date{\today}

\begin{abstract}
We describe a simple model, based on the preon model of Shupe 
and Harari, in which the binding of preons is represented topologically. 
We then demonstrate a direct correspondence between this model and 
much of the known phenomenology of the Standard Model. In particular
we identify the substructure of quarks, leptons and gauge bosons with 
elements of the braid group $B_3$. Importantly, the preonic objects of this 
model require fewer assumed properties than in the Shupe/Harari model, yet 
more emergent quantities, such as helicity, hypercharge, and so on, are found. 
Simple topological processes are identified with electroweak interactions and 
conservation laws. The objects which play the role of preons in this model
may occur as topological structures in a more comprehensive theory, and may 
themselves be viewed as composite, being formed of truly fundamental sub-components, 
representing exactly two levels of substructure within quarks and leptons. 
\end{abstract}

\keywords{preons \, composite models \, topology}
\pacs{12.60.Rc, 12.10.Dm}
\maketitle


\section{Introduction}
\label{sec:intro}
The Standard Model (SM) provides an extremely succesful and simple 
means of classifying and understanding the physical processes which fill the 
Universe. However the existence of many seemingly arbitrary features hints 
at a more fundamental physical theory from which the SM arises. Considering the 
successful series of ideas leading through molecules, to atoms, nuclei, nucleons, 
and quarks, it was perhaps inevitable that a model based on compositeness of 
quarks and leptons would be developed. The first such was 
proposed by Pati and Salam \cite{PatiSalam} in 1974, however it lacked any real 
explanatory power. Pati and Salam gave the name {\em preons} to their 
hypothetical constituent particles, and this name was gradually adopted to refer
to the sub-quark/sub-lepton particles of any model. Other notable preon models 
were developed by several authors (e.g. \cite{Akama}), but it is the 
so-called Rishon Model, proposed simultaneously by Harari and Shupe 
\cite{Harari,Shupe} which will be of most interest to us here. In Harari's more 
commonly quoted terminology, this model involves just two kinds of `rishons', one 
carrying an electric charge of $+e/3$ where $-e$ is the charge on the electron, 
the other neutral. The rishons combine into triplets, with the two 
``three-of-a-kind'' triplets being interpreted as the $\nu_e$ and $e^+$, and the 
permutations of triplets with an ``odd-man-out'' being interpreted as the different
colours of quarks. Equivalent combinations can be formed from the anti-rishons to 
create the remaining fermions and anti-fermions, such as the $e^-$ and 
$\anti{\nu_e}$. Certain combinations of rishons and anti-rishons were also 
suggested to correspond with gauge bosons. \\
The rishon model accounted for many aspects of the SM, including the precise ratios 
of lepton and quark electric charges, and the correspondence between fractional 
electric charge and colour charge. Unfortunately, as originally proposed it also 
had several problems, including the lack of a dynamical framework, and the lack of 
an explanation as to why the ordering of rishons within triplets should matter. 
A charge called ``hypercolour'' was proposed to solve these problems 
\cite{harari_seiberg}. The introduction of hypercolour implied the existence of 
``hypergluons'' and some QCD-like confinement mechanism for the rishons. Hence, 
the simplicity of the original model was reduced, and many of the fundamental 
questions about particles and interactions were simply moved to the realm of rishons,
yielding little obvious advantage over the SM. Furthermore preon models were never able 
to adequately answer several fundamental questions, such as how preons confined at all 
length scales experimentally probed can form very light composites (see e.g. 
\cite{mismatch} for an attempt to address this issue).\\ 
This article presents an idea based on the original rishon model (without 
hypercolour), which we call the Helon Model. The reader should note the subtle
yet important distinction that this is not a preon model {\em per se}, based upon 
point-like particles, but rather a preon-{\em inspired} model, which may be realised 
as a topological feature of some more comprehensive theory. For this reason we do not 
believe that the objections levelled at the rishon model and other preon models should 
be assumed, {\em a priori}, to be relevant to the helon model. A thorough investigation 
of such issues will be undertaken in subsequent work, however they are beyond the scope 
of the current article. Here we simply present a pedagogical introduction to the helon 
model, and describe how various features of the standard model emerge from it.

\section{The Helon model}
\label{sec:helons}
Let us now introduce our topologically-based toy model of quarks, leptons, and gauge bosons. 
It is convenient to represent the most fundamental objects in this model by twists 
through $\pm\pi$ in a ribbon. For convenience let us denote a twist through $\pi$ as a 
``dum'', and a twist through $-\pi$ as a ``dee'' ($U$ and $E$ for short, after 
Tweedle-dum and Tweedle-dee \cite{wonderland}). Generically we refer to such twists by 
the somewhat whimsical name ``tweedles''\cite{yesteryear}. We hope to deduce the 
properties of quarks and leptons and their interactions from the behaviour of 
their constituent tweedles, and to do so we shall employ a set of assumptions 
that govern their behaviour:\\
{\bf 1) Unordered pairing:} {\em Tweedles combine in pairs, so that their 
total twist is $\,0$ modulo $2\pi$, and the ordering of tweedles within a pair 
is unimportant}. The three possible combinations of $UU$, $EE$, and 
$UE \equiv EU$ can be represented as ribbons bearing twists through the 
angles $+2\pi$, $-2\pi$, and $0$ respectively. A twist through $\pm 2\pi$
is interpreted as an electric charge of $\pm e/3$. We shall refer to such pairs 
of tweedles as {\em helons} (evoking their helical structure) and denote the 
three types of helons by $H_+$, $H_-$, and $H_0$. \\ 
{\bf 2) Helons bind into triplets:} {\em Helons are bound into triplets by 
a mechanism which we represent as the tops of each strand being connected to each 
other, and the bottoms of each strand being similarly connected}. A triplet of 
helons may split in half, in which case a new connection forms at the top or bottom
of each resulting triplet. The reverse process may also occur when two triplets 
merge to form one triplet, in which case the connection at the top of one triplet 
and the bottom of the other triplet 
``annihilate'' each other. \\
The arrangement of three helons joined at the top and bottom is equivalent to two 
parallel disks connected by a triplet of strands. In the simplest case, such an 
arrangement is invariant under rotations through angles of $2\pi/3$, making it 
impossible to distinguish the strands without arbitrarily labelling or colouring 
them. However we can envisage the three strands crossing over or under each other 
to form a braid. The three strands can then be distinguished by their relative 
crossings. 
We will argue below that 
braided triplets represent fermions, while unbraided 
triplets provide the simplest way to represent gauge bosons.\\   
{\bf 3) No charge mixing:} {\em When constructing braided triplets, we will not 
allow $H_+$ and $H_-$ helons in the same triplet. $H_+$ and $H_0$ mixing, and 
$H_-$ and $H_0$ mixing are allowed}.\\
{\bf 4) Integer charge:} {\em All unbraided triplets must carry integer electric 
charge}.\\
Assumption 1) is unique to the helon model, as it reflects the composite nature
of helons, however assumptions 2), 3) and 4) are merely restatements in a different
context of assumptions that Shupe and Harari made in their work. We will require 
one further assumption, however it will be left until later, as it can be better 
understood in the proper context.\\
In terms of the number of fundamental objects, the helon model is more economical 
than even the rishon model, albeit at the cost of allowing helons to be composite. 
This seems a reasonable price to pay, as the tweedles are extremely simple, being 
defined by only a single property (i.e. whether they twist through $+\pi$ or 
$-\pi$). This may be interpreted as representing exactly two levels of substructure
within quarks and leptons.\\ 
The helon model also improves on the original rishon model by explaining why the 
ordering of helons (which are analogous to rishons) should matter. There are three 
permutations of any triplet with a single ``odd-man-out''. Without braiding we 
cannot distinguish these permutations, by the rotational invariance argument above.
However, if we allow braiding, the strands (and hence permutations) become distinct, 
in general. We may associate these permutations with the three colour charges of QCD, 
just as was done in the rishon model, and write the helons in ordered triplets for 
convenience. The possible combinations and their equivalent quarks are as follows
(subscripts denote colour):
\begin{center}
\begin{tabular}{llllll}
  $H_+H_+H_0$ & \hspace{-0.4mm}($u_B$)\hspace{1.2mm} 
     & $H_+H_0H_+$ & \hspace{-0.4mm}($u_G$) \hspace{1.2mm} & $H_0H_+H_+$ & \hspace{-0.4mm}($u_R$)\\
  $H_0H_0H_+$ & ($\anti{d_B}$) 
     & $H_0H_+H_0$ & ($\anti{d_G}$)
                       & $H_+H_0H_0$ & ($\anti{d_R}$)\\ 
  $H_-H_-H_0$ & ($\anti{u_B}$)
     & $H_-H_0H_-$ & ($\anti{u_G}$)
                       & $H_0H_-H_-$ & ($\anti{u_R}$)\\  
  $H_0H_0H_-$ & ($d_B$)  
     & $H_0H_-H_0$ & ($d_G$) & $H_-H_0H_0$ & ($d_R$).\\  
\end{tabular}
\end{center}
while the leptons are:
\begin{center}
\begin{tabular}{llllll}
$H_+H_+H_+$ & \hspace{-0.4mm}$(e^+)$ \hspace{1.2mm} 
   & $H_0H_0H_0$ & ($\nu_e$) \hspace{1.2mm}
     & $H_-H_-H_-$ & ($e^-$) \\
\end{tabular}
\end{center}
Note that in this scheme we have created neutrinos, but not anti-neutrinos.
This has occurred because the $H_0$ helon is its own anti-particle. This 
apparent problem will be turned to our advantage in Section~\ref{sec:phenom}. 
Note also that we have not ascribed any of the usual quantum numbers (such as mass or 
spin \footnote{The twist of tweedles may be viewed as a spin in some abstract space,
akin to isospin. We will not assume that this is the same as 
intrinsic angular momentum.}) to the helons. It is our expectation that spin, mass,
hypercharge and other quantities emerge dynamically, just like electric charge, as 
we arrange helons into more complex patterns.\\
It is interesting to note that three helons seems to be the minimum number 
from which a stable, non-trivial structure can be formed. By stable, we mean that 
a physical representation of a braid on three strands (e.g. made from strips of 
fabric) cannot in general be smoothly deformed into a simpler structure. By 
contrast, such a physical model with only two strands can always be untwisted.

\section{Phenomenology of the helon model}
\label{sec:phenom}
\begin{figure}[t]
\begin{center}
{\includegraphics[height=35mm,angle=0]{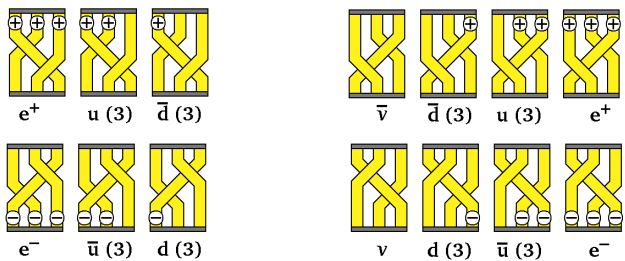}}\\
\end{center}
\caption{The fermions formed by adding zero, one, two or three charges to a neutral
braid. Charged fermions come in two handedness states each, while $\nu$ and 
$\anti{\nu}$ come in only one each. (3) denotes that there are three possible 
permutations, identified as the quark colours. The bands at top and bottom 
represent the binding of helons.}
\label{fig:all_fermion}
\end{figure}

\begin{figure}[t]
\begin{center}
{\includegraphics[height=25mm,angle=0]{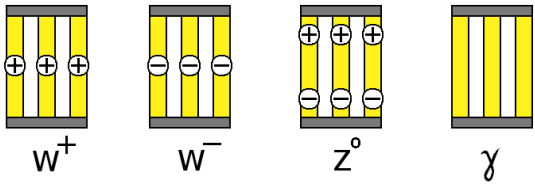}}
\end{center}
\caption{The bosons of the electroweak interaction. Notice that the $Z^0$ and 
the photon can deform into each other.}
\label{fig:bosons}
\end{figure}

\begin{figure}[t]
\begin{center}
{\includegraphics[height=32mm,angle=0]{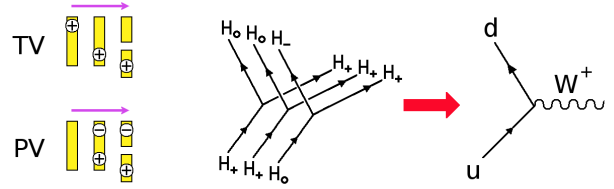}}
\end{center}
\caption{The charge transferring vertex (TV) and the polarising vertex 
(PV) are basic helon vertices from which the electroweak basic vertices 
can be formed.}
\label{fig:basic_vertex}
\end{figure}

A braid on $n$ strands is said to be an element of the braid group $B_n$. 
Given any element of $B_n$, it is also possible to create a corresponding 
anti-braid, which is the top-to-bottom mirror image of that braid, and is also 
an element of $B_n$ (it is therefore merely a matter of convention what we call 
a braid or an anti-braid). If we join the strands of a braid with those of its 
top-to-bottom mirror image (i.e. we take the braid product \cite{braid}) we 
obtain a structure which can be deformed into the trivial braid. Conversely, 
if the strands of a trivial braid are crossed so as to form a braid, an 
anti-braid must also be formed. Thus if we assign a quantity $\beta=+1$ to braids 
and $\beta=-1$ to anti-braids, $N_B=\sum\,\beta$ is conserved in splitting 
and joining operations (where the sum is taken over all braids and anti-braids present). 
These properties are reminiscent of the relationship between particles and anti-particles, 
and so it is natural to use the top-to-bottom mirroring of braids as a model for 
particle--anti-particle interchange, or C inversion. In addition, given any element 
of $B_n$, its left-to-right mirror image may be formed. We will call these the {\em left-handed} 
and {\em right-handed} forms of a braid. It seems natural to equate these with 
particles and anti-particles having positive and negative helicity. From the 
discussion above, all fermions and their anti-particles are represented by 
braids which are elements of $B_3$.\\
Let us construct the first-generation fermions with positive charge as shown on 
the right of Figure~\ref{fig:all_fermion} (we are using a basic braid, like that 
used to plait hair, for illustrative purposes, however an arbitrary number of 
crossings is possible). This yields the positron, up quark, anti-down 
quark, and anti-neutrino. Now let us construct the negatively-charged fermions by 
taking the top-to-bottom mirror images of the positively charged fermions. We have 
constructed the positively charged fermions by adding positive charges (dees) to a 
right-handed braid, and their anti-particles by adding negative charges (dums) to a
left-handed anti-braid. This decision was of course completely arbitrary. We can 
also add dees to a left-handed braid and dums to a right-handed anti-braid. If we 
do this, we create all the possible charge-carrying braids in two different 
handedness states exactly once, but following the same procedure for the uncharged 
braids (i.e. neutrinos and anti-neutrinos) would mean duplicating them, since this 
second pair of neutral leptons is identical to the first pair, rotated through 
$\pm\pi$. In other words, to avoid double-counting we can only construct the 
(anti-)neutrino in a (right-)left-handed form, while all the other fermions come 
in both left- and right-handed forms. This pleasing result is a direct consequence 
of the fact that we construct the neutrino and anti-neutrino from the same 
sub-components (by contrast the rishon model used neutral rishons for the $\nu_e$ 
and neutral anti-rishons for the $\anti{\nu_e}$).\\
If we perform C and P operations on any braid (except a $\nu$ or $\anti{\nu}$, 
on which we cannot perform P) we obtain the braid diagonally opposite it in 
Figure~\ref{fig:all_fermion}. We may define a further operation which consists 
of rotating a braid clockwise or anti-clockwise through $\pi$, and reversing 
the sign of all charges. This operation will be called T, and we note that 
performing C, P, and T in any order on 
a braid leaves that braid unchanged.\\
Having constructed the quarks and leptons, we now turn our attention to their 
interactions via the electroweak and colour forces.\\ 
{\bf{\em The Electroweak Interaction}}
We shall begin by constructing the bosons of the electroweak interaction, $\gamma$,
$W^+$, $W^-$, and $Z^0$. The $W^+$ and $W^-$ may be regarded as a triplet of $H_+$s
and a triplet of $H_-$s respectively. We can create neutral bosons from a triplet 
of similar helons in two ways. One is as a triplet of untwisted helons, the other 
as a triplet of ``counter-twisted'' helons (that is, each helon carries explicit 
left-handed and right-handed twists). We shall claim that the former is the photon,
the latter is the $Z^0$ (we may also speculate that deforming an untwisted 
helon into a counter-twisted helon, or vice-versa, accounts for the Weinberg mixing 
\begin{figure}[t]
\begin{center}
{\includegraphics[height=35mm,angle=0]{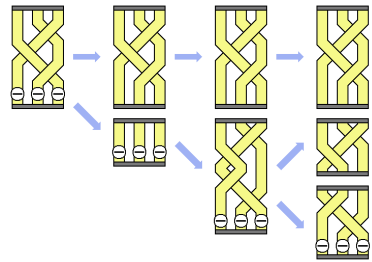}}\\
\end{center}
\caption{A representation of the decay $\mu \rightarrow \nu_\mu + e^- + \anti{\nu_e}$, 
showing how the substructure of fermions and bosons demands that charged leptons decay
to neutrinos of the same generation.}
\label{fig:muon}
\end{figure}
between the $Z^0$ and the photon \footnote{To be consistent with the GWS model we should 
probably say that these braids represent the $W^0$ and $B$. Conceptually there is 
little difference either way.}). What sets bosons apart from fermions (i.e. so that 
a $\gamma$ is distinct from a neutrino, and a $W^\pm$ is distinct from an $e^\pm$) 
is that the strand permutation induced by the braid that forms a boson is the 
identity permutation. The simplest braid that fulfills this criterion is the 
trivial braid, as in Fig.~\ref{fig:bosons}.\\
All interactions between helons can be viewed as cutting or joining operations, 
in which twists (tweedles) may be exchanged between helons. These operations define
basic vertices for helon interactions. By combining three of these basic helon vertices
in parallel we construct the basic vertices of the electroweak 
interaction (Figure~\ref{fig:basic_vertex}). Crossing symmetries of the helon 
vertices automatically imply the usual crossing symmetries for the electroweak 
vertices.\\ 
We can represent 
higher generation fermions by allowing the three helons to cross in more 
complicated patterns. Thus all fermions may be viewed as a basic neutrino 
``framework'' to which electric charges are added, and the structure of this 
framework determines to which generation a given fermion belongs. If, say, a 
muon emits charge in the form of a $W^-$ boson, the emission of this (trivially 
braided) boson does not affect the fermion's braiding structure. Therefore the 
muon loses charge but does not change generation, and must transform into a 
muon-neutrino. Likewise, tau-neutrinos may only transform into tau particles and 
electron-neutrinos may only transform into electrons. A similar argument applies 
to quarks, although the mechanism by which the Cabibbo angles become non-zero is 
not immediately apparent. Notice that since $W^\pm$ bosons carry 
three like charges, such processes will 
change quark flavour but not colour.\\
{\bf {\em The Colour Interaction}}
We have thus far described electroweak interactions as splitting and joining 
operations on braids. We may similarly represent colour interactions physically, 
this time as the formation of a ``pancake stack'' of braids. Each set of strands 
that lie one-above-the-other can be regarded as a ``super-strand'', with a total 
charge equal to the sum of the charges on each of its component strands. If we 
represent braids as permutation matrices, with each non-zero component being a 
helon, we can easily represent the colour interaction between fermions as the 
sum of the corresponding matrices. For instance a red up quark, and an anti-red 
anti-up combining to form a neutral pion could be represented as 
\begin{eqnarray*}
 \left[\begin{array}{ccc}
   0  &  0  & H_0   \\
  H_+ &  0  & 0   \\
   0  & H_+ & 0  
      \end{array}\right] 
& + & \left[\begin{array}{ccc}
   0  &  0  & H_0   \\
  H_- &  0  & 0   \\
   0  & H_- & 0  
      \end{array}\right] \hspace{8mm} \\ 
\end{eqnarray*}
\vspace{-9mm}
\beqa
\hspace{7mm}& = & \left[\begin{array}{ccc}
   0        &  0        & H_0 + H_0   \\
  H_+ + H_- &  0        & 0   \\
   0        & H_+ + H_- & 0  
      \end{array}\right] 
\eeqa
where the sum of two helons on the right-hand-side denotes a super-strand composed 
of a pair of helons. In this case all three super-strands have zero net charge. 
In this way hadrons can be regarded as a kind of superposition of quarks. 
We can now introduce our last assumption.\\
{\bf 5) Charge equality:}{\em When two or more braids are stacked, 
they must combine to produce the same total charge on all 
three super-strands}.\\
It is obvious that leptons trivially fulfill this criterion, and will 
therefore not partake of the colour interaction. The reader may verify that 
quarks will naturally form groupings with integer electric charge, which are 
colour-neutral, for example three quarks forming a proton:
\begin{eqnarray*} 
d_B+u_R+u_G & =  &  \left[\begin{array}{ccc}
   0  &  0  & H_0   \\
  H_0 &  0  & 0   \\
   0  & H_- & 0  
      \end{array}\right] \hspace{24mm}\\
\end{eqnarray*}
\vspace{-9mm}
\beqa
\hspace{7mm} & + & \left[\begin{array}{ccc}
                               0   &  0  & H_0   \\
			       H_+ &  0  & 0   \\
			       0   & H_+ & 0 
                               \end{array}\right] +
 \left[\begin{array}{ccc}
   0  &  0  & H_+   \\
  H_0 &  0  & 0   \\
   0  & H_+ & 0 
      \end{array}\right].
\eeqa
In QCD there are eight gluons. Two of these are superpositions of like colour/anti-colour states, which we 
may visualise as triplets with some untwisted strands, and some counter-twisted strands
(strands with explicit positive and negative twist, as per the description of the 
$Z^0$, above). The remaining six are ``pure'' unlike colour/anti-colour pairings, which 
we can visualise as the six permutations of one $H_+$, one $H_0$, and one $H_-$.\\     
{\bf{\em Hypercharge}}
Recall that we assigned $\beta=+1$ to the braids on the top row of 
Figure~\ref{fig:all_fermion}, and $\beta=-1$ to the braids 
on the bottom row. This effectively distinguishes between fermions with a net positive 
and net negative charge. To distinguish between quarks and leptons the number of 
odd-men-out within any triplet, divided by the number of helons within a triplet 
suggests itself as a useful quantity (taking a value of $0$ for leptons and $1/3$ 
for quarks). However this quantity does not distinguish between a particle and its 
anti-particle. To rectify this shortcoming we will define a new quantity, given 
by one-third the number of ``more positive'' helons, minus one-third the number 
of ``less positive'' helons. We shall denote this quantity by the symbol $\Omega$. 
To clarify, $H_+$ helons are considered ``more positive'' than $H_0$ helons, which 
are ``more positive'' than $H_-$ helons. If $N(H_+)$ is the number of $H_+$ 
helons, $N(H_0)$ the number of $H_0$s and $N(H_-)$ the number of $H_-$s within a 
triplet, and remembering that $H_+$ and $H_-$ helons never occur within 
the same braided triplet, we may write
\beq
\Omega=\beta\left(\frac{1}{3}N(H_+)+\frac{1}{3}N(H_-)-\frac{1}{3}N(H_0)\right).
\label{eq:Omega}
\eeq
Hence we have $\Omega=+1$ for the $e^+$, $\Omega=+1/3$ for the $u$, $\Omega=-1/3$ 
for the $\anti{d}$, and $\Omega=-1$ for the anti-neutrino. For the electron, 
anti-up, down, and neutrino the signs are reversed. With this definition,  
noting that $N(H_0)=3-(N(H_+)+N(H_-))$ and the total electric charge of a fermion 
is given by 
\beq
Q = \beta\left(\frac{1}{3}N(H_+)+\frac{1}{3}N(H_-)\right),
\eeq  
it is easy to show that 
\beq
Q = \frac{1}{2}\left(\beta + \Omega\right).
\label{eq:QbetaOmega}
\eeq
The values of $\Omega$ and $\beta$ for the second and third-generation fermions 
are the same as those of their first-generation counterparts. For the quarks and 
anti-quarks $\Omega$ reproduces the SM values of strong hypercharge, while for 
the leptons $\beta$ reproduces the SM values of weak hypercharge. Within the SM, 
strong hypercharge is defined in terms of baryon number, strangeness, charm, 
bottomness and topness as $Y=A+S+C+B+T$. In the helon model $\Omega$ plays a role 
equivalent to baryon number, thus $Y = \Omega$ for quarks, and there are no 
analogues of $S$, $C$, $B$, or $T$. This should be viewed not as a deficiency, 
but as a simplification, reminiscent of that recently proposed by Robson 
\cite{Robson}. Similarly, within the SM weak hypercharge for leptons is defined 
in terms of electron number, muon number, and tau number as 
$y = -(L_e + L_\mu + L_\tau)$. In the helon model $\beta$ plays a role equivalent
to lepton number, thus $y = \beta$ for leptons. We also observe that for quarks 
$\beta/2$ reproduces the values of the third component of strong isospin, while 
for leptons $\Omega/2$ reproduces the values of the third component of weak isospin 
(in short, the roles of $\beta$ and $\Omega$ as isospin and hypercharge are 
reversed for quarks and leptons). With these correspondences the 
Gell-Mann--Nishijima relation $Q=I_3+Y/2$ for quarks may trivially be derived from 
Eq.~(\ref{eq:QbetaOmega})\\
It should be noted that in addition to the absence of analogues of $S$, $C$, $B$, 
or $T$, in the helon model the second and third-generation quarks carry non-zero
values of strong isospin. Hence the $c$-$s$ and $t$-$b$ pairs appear to be strong 
isospin doublets, like the $u$-$d$ pair. By contrast, in the SM this view is 
rejected, and the second and third-generation quarks are assigned $I_3=0$ due to
the large differences in mass between the $c$ and $s$ quarks, and between the $t$ 
and $b$ quarks. We shall make no further comment on this point, other than to 
suggest that until a reliable way of calculating the quark masses from first 
principles is found, we cannot exclude the possibility that the percieved connection 
between mass and isospin (based initally on the example of the proton and neutron, 
which are now known to be composite) is incomplete.\\  
{\bf{\em Conservation of lepton and baryon number}}
As we noted earlier braids are conserved, that is, braids and anti-braids are
created and destroyed together, from trivial braids (i.e. bosons). This allows for 
pair annihilation and creation of 
fermions within the helon model. \\
Let $x$ and $y$ represent triplets of helons, and let $N_{\mathrm{B}}(x)$ and 
$N_{\anti{\mathrm{B}}}(y)$ be, respectively, the number of braids composed of the 
triplet $x$ and anti-braids composed of the triplet $y$ (e.g. 
$N_{\anti{\mathrm{B}}}(H_-H_-H_-)$ is the number of anti-braids composed of three 
$H_-$ helons). Furthermore let $N(f)$ be the number of, and $\beta(f)$ be the value 
of $\beta$ associated with, fermions of type $f$. Then, for an arbitrary collection 
of leptons, the total lepton number is 
\vspace{-1mm}
\beqa 
L & = & \beta(e^+)N(e^+)+ \beta(e^-)N(e^-) \nonumber \\
  &   &  + \beta(\anti{\nu_e})N(\anti{\nu_e})+
        \beta(\nu_e)N(\nu_e) \nonumber\\
  & = & N_{\mathrm{B}}(H_+H_+H_+)+N_{\mathrm{B}}(H_0H_0H_0) \nonumber \\
  &   & -N_{\anti{\mathrm{B}}}(H_-H_-H_-)-N_{\anti{\mathrm{B}}}(H_0H_0H_0)
	                                                     \nonumber \\
  & = & N_\mathrm{B} - N_{\anti{\mathrm{B}}} \nonumber \\  
  & = & N_\mathrm{B}^{\mathrm{Total}} 
  \eeqa
and so conservation of braids implies conservation of lepton number. This 
argument can easily be refined to account for the seperate conservation of 
electron number, muon number, and tau number. A similar argument applies to 
quarks and baryon number.

\section{Unresolved issues}
\label{sec:issues}
There are three significant issues the helon model has not yet 
addressed: the origin of spin, the origin of mass, and the nature 
of Cabbibo-mixing.\\ 
At the outset we hoped that spin would be emergent. Thus far we have been partly 
successful, identifying helicity of helon triplets, but not spin itself. It may 
be tempting then to assume that helons carry spin-$\frac{1}{2}$, and combine to 
form spin-$\frac{1}{2}$ composites. However this would not explain what spin is, 
would complicate the model, and would necessitate further assumptions to explain 
why helons never combine to form spin-$\frac{3}{2}$ states. It therefore seems 
reasonable to press ahead in the hope that spin will emerge as a consequence, 
rather than an assumption, of a dynamical 
theory based on the helon model.\\
We can speculate briefly on the origin of mass by noting that the helons within a 
fermion are represented as crossing each other to form structures which are elements 
of $B_3$. Similarly, when a helon is twisted, its left and right edges cross each 
other, forming an object equivalent to a generator of $B_2$. Therefore twisting 
(electric charge) and braiding are fundamentally the same process
. Since the simplest fermions (neutrinos) are also the least 
massive fermions, and more complex braiding patterns correspond with higher 
generations (and hence greater mass) we propose that, roughly speaking, the more 
non-trivial the crossings in a fermion's substructure, the greater its mass. Much 
work will be needed to develop and test this idea quantitatively. Connected to the 
issue of mass is that of gravity. We note that our construction of the gauge bosons 
did not include a helon triplet that corresponds to the graviton (or for that 
matter, the Higgs). However gravity and the helon model may arise as seperate parts of some
more comprehensive theory (see Sec.~\ref{sec:discussion} for further discussion of this 
possibility).\\
Cabbibo-mixing may be related to the presence of $H_0$ helons within quarks. 
This would lead to an obvious connection between Cabbibo-mixing and neutrino 
oscillations, and explain why charged leptons do not mix between generations.
A modification of the helon model in which the presence of $H_0$ helons within 
a triplet enable inter-generational mixing therefore seems like a reasonable 
avenue of future enquiry.

\section{Discussion}
\label{sec:discussion}
The Helon Model differs from previous preon models in that we have chosen to 
describe the properties of composite fermions and bosons not in terms of 
fundamental properties of the preons themselves (e.g. hypercolour), but in the 
interrelations between the preons (in essence extending the idea behind 
Harari and Shupe's original explanation of colour charge to other quantitiess). 
We note once again that the helon model should, to be precise, not even be viewed 
as a preon model, but as a preon-inspired set of mathematical properties.
The significance of this model is its extreme economy and extension of earlier 
ideas. The rishon model explained the number of leptons and quarks, the precise
ratios of their electric charges, and the origin and nature of colour charge.
The helon model does all this, but in addition it does away with the {\em ad hoc} 
assumption the permutations of preonic objects within a triplet are distinct, and 
explains why the the ordering of should matter (leading to three distinct colour 
charges) in a natural way. A secondary consequence of this is to explain why charged 
fermions come in two varieties of helicity, while neutrinos come in only one. The 
helon model shows that baryon number (of quarks, not hadrons) and lepton number 
should be conserved in all weak interactions. We have derived the 
Gell-Mann--Nishijima relation from first principles using a simple counting argument.
The helon model provides an explanation for the existence of generations, which 
predicts that neutrinos of a given generation will only interact directly with 
charged leptons of the same generation. All this is achieved using a much simpler 
set of ``particles'' and ``interactions'' than those of the rishon model - namely 
one type of fundamental object, a crossing operation, and a splitting/joining 
process.\\
Rather than being viewed as distinct types of processes, decays which violate 
strangeness, charm, bottomness and topness can be classified together as simply 
'generation-changing decays' since $S$, $C$, $B$, and $T$ have been demoted from 
their presumed roles as valid quantum numbers.\\ 
By representing the braided substructures of fermions and bosons as permutation 
matrices, we represented colour interactions as matrix addition. Interestingly the 
electroweak interactions behave as matrix multiplication. This point shall be taken 
up in more detail in further work, currently in preparation.\\     
While the helon model explains many things, there are a number of issues it raises. 
Foremost among these is the question of just what physical process (if any) the 
twisting and braiding of helons represents. The correspondece between the topologcal 
structures of the helon model and the fermions and bosons of the standard model may 
represent a new way of interpreting states that occur in topological theories, such as 
M-theory. It was recently proposed that the helons may represent connection variables 
in the framework of Loop Quantum Gravity \cite{Smolin}. Development
of this idea is reported in \cite{SundanceFotiniLee}. This represents the first step 
towards developing a dynamical theory of all interactions, particles, and spacetime 
itself, incorporating the helon model. Further investigation is required to see whether 
the helon model is consistent with all observed particle interactions, and whether it 
predicts interactions that are not observed. These and related issues are sure to 
determine the directions of future research. 

\section{Acknowledgements}
The author is extremely grateful to Alex Kalloniatis for numerous helpful comments 
on style and content, Larisa Lindsay for proofreading, and to L. J. Nickisch and 
John Hedditch for encouragement and helpful discussions.
  

\end{document}